\documentclass[prl,twocolumn,floatfix,superscriptaddress]{revtex4-1}
\usepackage{dcolumn,amsmath}
\usepackage{graphicx}
\usepackage{bm}
\usepackage{hyperref}

\newcommand{\Eeff}{\ensuremath{E_{\rm eff}}}
\newcommand{\eEDM}{{\em e}EDM}
\newcommand{\ecm}{\ensuremath{e {\cdotp} {\rm cm}}}
\newcommand{\cm}{\ensuremath{{\rm cm}}$^{-1}$}
\begin{document}
\title{Theoretical study of ThO for the electron electric dipole moment search}

\author{L.V.\ Skripnikov}\email{leonidos239@gmail.com}
\author{A.N.\ Petrov}
\author{A.V.\ Titov}
\homepage{http://www.qchem.pnpi.spb.ru}
\affiliation{Federal state budgetary institute ``Petersburg Nuclear Physics Institute'', Gatchina, Leningrad district 188300, Russia}
\affiliation{Dept.\ of Physics, Saint Petersburg State University, Saint Petersburg, Petrodvoretz 198904, Russia}
 \date{\today}

\begin{abstract}
An experiment to search for the electron electric dipole moment (\eEDM) on the metastable $H^3\Delta_1$ state of ThO molecule was proposed and now in the final stage of preparation by the ACME collaboration [http://www.electronedm.org].  
To interpret the experiment in terms of \eEDM\ and dimensionless constant $k_{T,P}$
characterizing the strength of the scalar T,P-odd electron-nucleus neutral current interaction,
an accurate theoretical study of effective electric field on electron, Eeff,
and $W_{T,P}$ constants is required.
We report calculation of
\Eeff\ (84~GV/cm) and a parameter of T,P-odd scalar neutral currents interaction, $W_{T,P}$
(116 kHz), together with the  hyperfine structure constant, molecule frame dipole moment and $H^3\Delta_1\to X^1\Sigma^+$ transition energy, which can serve as a measure of reliability of the obtained \Eeff\ and $W_{T,P}$ values.
Besides, our results include a parity assignment and evaluation of the electric-field dependence for the magnetic $g$ factors for the $\Omega$-doublets of $H^3\Delta_1$.
\end{abstract}

\maketitle


One of the most intriguing fundamental problems of modern physics is the search for a permanent electric dipole moment (EDM) of elementary particles. A nonzero value of EDMs implies manifestation of interactions which are not symmetric with respect to both time (T) and spatial (P) inversions (T,P-odd interactions). Particularly, the observation of electron EDM (\eEDM) at the level significantly greater than $10^{-38}$ would indicate the presence of a ``new physics'' beyond the Standard model. Popular extensions of the Standard model predict the magnitude of the \eEDM\ at the level of $10^{-26}-10^{-28}$ \cite{Commins:98}. The most rigid upper bound on the \eEDM\ is attained in the experiments on a beam of YbF molecular radicals~\cite{Hudson:11a} ($1.05 \cdotp 10^{-27}\ecm $) and in the measurements with atomic Tl beam~\cite{Regan:02} ($1.6 \cdotp 10^{-27}\ecm$).

Nowadays a number of other prospective experiments are suggested and in part prepared \citep{Leanhardt:2011, Meyer:08, Cossel:12} which promise to achieve a sensitivity to \eEDM\ up to $10^{-29}-10^{-30}\ecm$. 
One of the most promising experiments towards the measurement of \eEDM\ is proposed and
now  prepared on the metastable $^3\Delta_1$ state of the thorium monoxide (ThO) molecule by ACME collaboration (groups of DeMille, Gabrielse, and Doyle) \cite{Vutha:2010, Vutha:2011}.
A very high sensitivity to \eEDM\ is expected in the nearest future, up to an order of magnitude and more than that attained in the YbF and Tl experiments, due to some unique combination of experimental advantages of the molecule. Even the value for \eEDM\ compatible with zero will lead to serious consequences for the modern theory of fundamental symmetries.

To interpret the measured data in terms of the \eEDM\ one should know a parameter usually called ``the effective electric field on electron'', \Eeff, which cannot be measured. To obtain \Eeff\ theoretically one can evaluate an expectation value of some T,P-odd operator (discussed in Refs.\ \cite{Kozlov:87, Kozlov:95, Titov:06amin}):
\begin{equation}
\label{matrelem}
W_d = \frac{1}{\Omega}
\langle \Psi|\sum_i\frac{H_d(i)}{d_e}|\Psi
\rangle,
\end{equation}
where $\Psi$ is the wave function of the considered state, and
$\Omega= \langle\Psi|\bm{J}\cdot\bm{n}|\Psi\rangle$,
$\bm{J}$ is the total electronic momentum, $\bm{n}$ is the unit vector along
the molecular axis directed from Th to O,
\begin{eqnarray}
  H_d=2d_e
  \left(\begin{array}{cc}
  0 & 0 \\
  0 & \bm{\sigma E} \\
  \end{array}\right)\ ,
 \label{Wd}
\end{eqnarray}
$\bm{E}$ is the inner molecular electric field, and $\bm{\sigma}$ are the Pauli matrices. In these designations $E_{\rm eff}=W_d|\Omega|$.
 
In addition to the interaction given by operator (\ref{Wd}) there is a scalar
T,P-odd electron-nucleus neutral currents interaction with the dimensionless constant 
$k_{T,P}$. The interaction is given by the following operator (see \cite{Hunter:91}):
\begin{eqnarray}
  H_{T,P}=i\frac{G\alpha}{\sqrt{2}}Zk_{T,P}\gamma_0\gamma_5n(\textbf{r}),
 \label{Htp}
\end{eqnarray}
where G is the Fermi constant, $\gamma_0$ and $\gamma_5$ are the Dirac matrixes and $n(\textbf{r})$ is the nuclear density normalized to unity.
To extract the fundamental $k_{T,P}$ constant from an experiment one need to know the factor $W_{T,P}$ that is determined by the electronic structure of a studied molecule on a given nucleus:
\begin{equation}
\label{WTP}
W_{T,P} = \frac{1}{\Omega}
\langle \Psi|\sum_i\frac{H_{T,P}(i)}{k_{T,P}}|\Psi
\rangle.
\end{equation}
Similarly to \Eeff\ parameters $W_{T,P}$ cannot be measured experimentally
and have to be obtained from the molecular electronic structure calculations.

A commonly used way of verification the theoretical \Eeff\ and $W_{T,P}$ values is to calculate ``on equal footing'' (the same approximation for the wave function) those molecular characteristics (properties or effective Hamiltonian parameters) which have comparable to 
\Eeff\ and $W_{T,P}$ sensitivity to different variations of wave function but, in contrast, can be measured. Similar to \Eeff\ and $W_{T,P}$ these parameters should be sensitive to a change of densities of the \textit{valence} electrons in atomic cores. The hyperfine structure constant, $A_{||}$, is traditionally used as such a parameter (e.g., see \cite{Kozlov:97}) and this is a valid touchstone for the ThO case as well. To obtain $A_{||}$ on Th theoretically, one 
can evaluate the following matrix element:
\begin{equation}
 \label{Apar}
A_{||}=\frac{\mu_{\rm Th}}{I\Omega}
   \langle
   \Psi|\sum_i\left(\frac{\bm{\alpha}_i\times
\bm{r}_i}{r_i^3}\right)
_z|\Psi
   \rangle, \\
\end{equation}
where $\mu_{\rm Th}$ is a magnetic moment of Th nucleus having spin $I$.

To validate our present study of the \Eeff, $W_{T,P}$ and the hyperfine structure constant $A_{||}$ for the $^3\Delta_1$ state of $^{229}$ThO, we have also performed calculations of the $H^3\Delta_1\to X^1\Sigma$ transition energy and the molecule-frame dipole moment.

 \section{Theoretical details}

The evaluation of \Eeff, $W_{T,P}$ and $A_{||}$ is usually a challenging problem for modern ab initio methods when studying systems containing heavy transition metals, lanthanides and, particularly, actinides (such as Th in the present consideration). An accurate theoretical investigation of such systems should take account of both the relativistic and correlation effects with the best to-date accuracy. It follows from Eqs.~(\ref{Wd})--(\ref{Apar})
that the operators related to \Eeff, $W_{T,P}$ and $A_{||}$ are essentially localized in the atomic core region. On the other hand the main contribution to the corresponding matrix elements is due to the valence electrons since contributions from the closed inner-core shells compensate each other in most cases of practical interest  for the operators dependent on the total angular momentum and spin.
It was shown by our group (see \cite{Titov:06amin} and references) that the problem of computation of such characteristics can be significantly simplified by splitting the calculation on two steps.
At the first step the electron correlation for valence (and outer-core electrons for better accuracy) is taken into account in a molecular calculation using some method of electron correlation treatment such as (coupled clusters, contiguration interaction, etc.), whereas the core (inner-core) electrons are excluded from this calculation using the generalized relativistic effective core potential (GRECP) \cite{Mosyagin:10a, Titov:99}, which yields an accurate valence region wave function by the most economical way. Secondly, since the inner-core parts of the valence one-electron ``pseudo-wavefunctions'' are smoothed within the GRECP method, they have to be recovered using some core-restoration method \cite{Titov:06amin}. The non-variational restoration is based on a proportionality (scaling) of valence and virtual spinors in the inner-core region of heavy atoms (e.g., see \cite{Titov:99} for details). The two-step approach has been recently used in \cite{Skripnikov:09, Skripnikov:11a, Petrov:11, Lee:13a, Petrov:13} for calculation of a number of characteristics, such as hyperfine structure constants, electron electric dipole moment enhancement factor, etc., in molecules and atoms. Besides it has been extended to the case of crystals in Ref.~\cite{Skripnikov:13b}.

Recently \cite{Skripnikov:13b} we have developed a code of nonvariational restoration which has been interfaced to {\sc dirac12} \cite{DIRAC12} and {\sc mrcc} \cite{MRCC2013}
 codes. These codes are used in the present paper. Scalar-relativistic calculations (i.e.\ without spin-orbit terms in the GRECP operator) were performed using {\sc cfour} code \cite{CFOUR}.

\section{Results and discussions}

The $1s-4f$ inner-core electrons of Th were excluded from molecular correlation calculations using the valence (semi-local) version of GRECP \cite{Mosyagin:10a} operator. Thus, the outermore 38 electrons were treated explicitly. Basis set for Th was constructed using the generalized correlated scheme \cite{Mosyagin:00}. It consists of $30s$, $8p$, $6d$, $4f$, $4g$ and $1h$ contracted Gaussians \footnote{
  $s$-type basis functions of Th are uncontracted.
}.
For oxygen the aug-cc-pvqz basis set \cite{Kendall:92} reduced to $6s$, $5p$, $4d$ and $3f$ contracted Gaussians was employed.

According to experimental data \cite{Huber:79} the internuclear distances for the ground $^1\Sigma^+$ and excited $^3\Delta_1$ states of ThO are about 3.5 a.u. Therefore, calculations of the states were performed with the given distance.
Calculations of the transition energy between these states as well as the molecule-frame dipole moment, \Eeff, $W_{T,P}$ and $A_{||}$ constants for the $^3\Delta_1$ state of ThO were performed using the single reference two-component relativistic coupled clusters with single, double and perturbative treatment of triple cluster amplitudes (CCSD(T))
 \footnote{For both the $^1\Sigma^+$ and $^3\Delta_1$ states the same set of one-electron spinors obtained in the two-component Hartree-Fock calculation were employed. Particularly, the reference determinant for the single-reference coupled clusters calculation of $^3\Delta_1$ state was constructed from this set of spinors. Such choice of one-electron functions is justified for the case of treatment by the coupled clusters approaches which include single clusters amplitudes. The latter are required to account for relaxation effects (see e.g. \cite{Bartlett:95}).}.
In addition, the basis set enlargement corrections to the considered parameters were also calculated. For this we have performed: (i) scalar-relativistic CCSD(T) calculation using the same basis set as used for the two-component calculation; (ii) scalar-relativistic CCSD(T) calculation using extended basis set on Th (with added $f$, $g$, $h$ and $i$ Gaussians). Corrections were estimated as a difference between the values of the corresponding parameters. The results are given in Table~\ref{TResults}.

\begin{table}[!h]
\caption{
The calculated values of transition energy ($T_e$), molecule-frame dipole moment ($d$), effective electric field (\Eeff), parameter of the T,P-odd scalar neutral currents interaction ($W_{T,P}$) and hyperfine structure constant (A$_{||}$) using the coupled clusters methods.
}
\label{TResults}
\begin{tabular}{ l  c  c  c  c  c}
\hline\hline
Method & $T_e,$     & $d$,  & \Eeff, & $W_{T,P}$, & A$_{||}$,  \\
       & $cm^{-1}$  & Debye & GV/cm & kHz & $\frac{\mu_{\rm Th}}{\mu_{\rm N}}\cdot$MHz \\
\hline
  2c-CCSD      & {} 5443 & {} 4.22 & {} 87 & 118  & -2953 \\  
  2c-CCSD(T)   & {} 6054 & {} 4.17 & {} 84 & 116  & -2880 \\ 
\hline  
  2c-CCSD(T)   & {} 5741 & {} 4.27 & {} 84 & 116 & --- \\ + basis corr.  & {}      & {}      & {}    &     \\
  
  Experiment  
\cite{Huber:79}  
      & {} 5321 & {} 4.24 $\pm$ 0.1 & {} ---& {} --- & --- \\ 
\hline\hline
\end{tabular}
\end{table}

The calculated value of transition energy is in a very good agreement with experimental datum, the  deviation, 420~\cm, is on the level of accuracy early attained by our group for compounds of transition metals and lanthanides.

It was recently shown in \cite{Safronova:13} that the magnetic moment of $^{229}$Th nucleus determined earlier \cite{Gerstenkorn:74} is inaccurate. Therefore, A$_{||}$ is given
in Table~\ref{TResults} in the units of $\mu_{\rm Th} / \mu_{\rm N} \cdot$MHz (where $\mu_{\rm N}$ is the nuclear magneton) in Table~\ref{TResults} to exclude the uncertainty of $\mu_{\rm Th}$ from our result. One can see from Table~\ref{TResults} that a good convergence of \Eeff\ with respect to both the basis set enlargement and correlation level is achieved. Taking into account the results from table \ref{TResults} as well as our earlier studies within the two-step procedure (e.g., see \cite{Lee:13a}) with calculating the 
\Eeff, $W_{T,P}$ and A$_{||}$ we expect that the theoretical uncertainty for our final values of 
the constants is smaller than 15\%. Unfortunately, there are no experimental data on A$_{||}$ up to now. Therefore, the corresponding indirect experimental verification of accuracy of \Eeff\ (see above) can not be performed to-date and further experimental measurements of A$_{||}$ are required.

%
%

\begin{figure}[pH]
\includegraphics[width = 3.3 in]{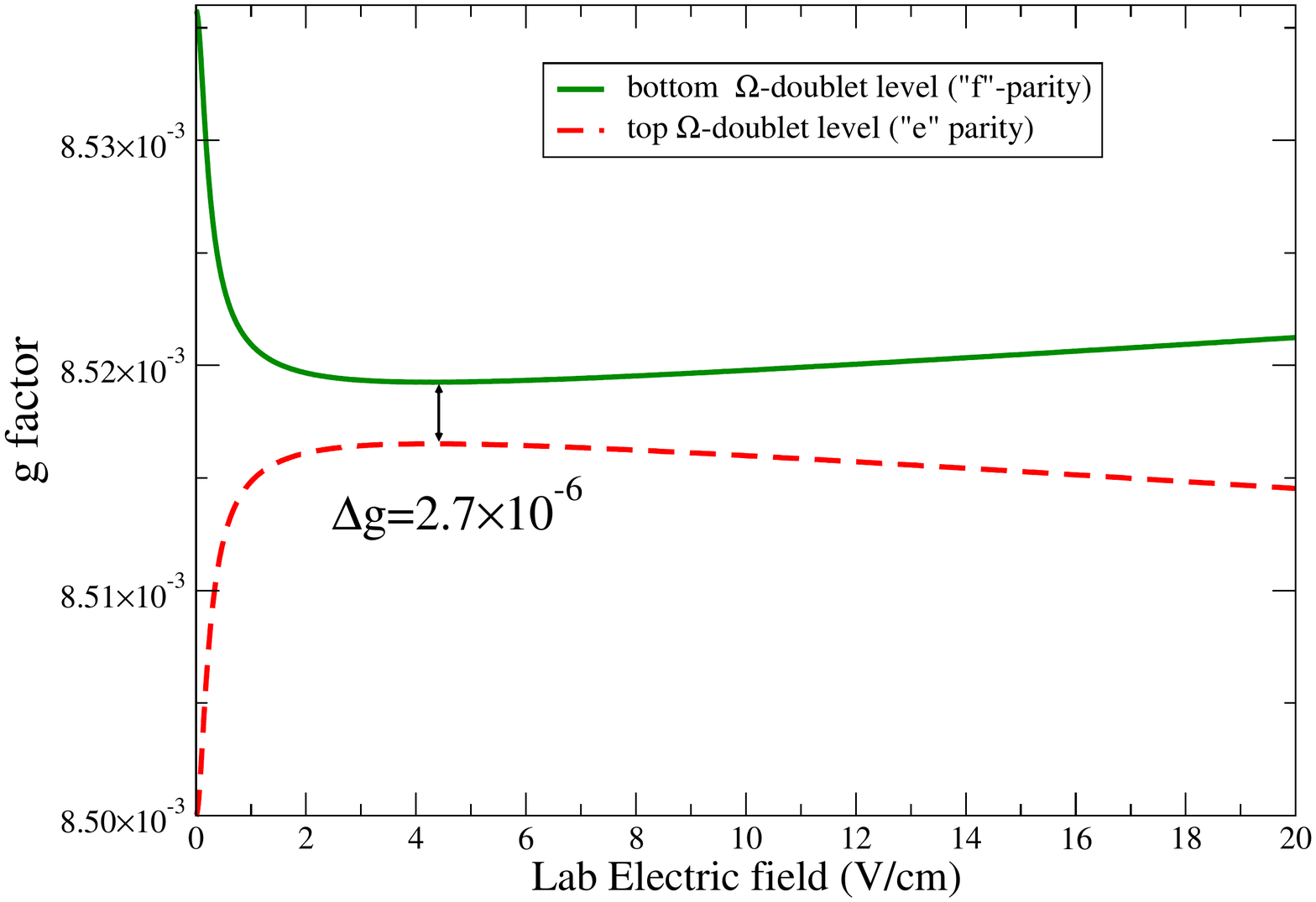}
 \caption{(Color online) Calculated $g$-factor curves for $J=1$ rotational level of $^{232}$Th$^{16}$O.} 
 \label{gfgecross}
\end{figure}

In the \eEDM\ search experiment on the ThO molecule, the \eEDM\ induced Stark splitting between the $J = 1,M = \pm1$ states of $e$ (parity is $(-1)^J$) or $f$ (parity is $-(-1)^J$) levels of the $\Omega-$doublet is measured. The $H^3\Delta_1$ state has a very small magnetic moment, $\mu_{H[ThO]}=8.5(5)\times 10^{-3} \mu_B$ \cite{Vutha:2011}, where $\mu_B$ is the Bohr magneton. The latter is a benefit for suppressing systematic effects due to spurious magnetic field. In a polarized molecule the $e$ and $f$ levels have opposite signs of \Eeff\ and almost identical $g$ factors. Therefore, when taking the difference between the splitting for $e$ and $f$ levels further suppression of the systematics is possible \cite{DeMille:2001}. The small difference between $g$ factors, $\Delta g$, comes from interactions of $H^3\Delta_1$ with $0^+$ and $0^-$ electronic states \cite{Petrov:11}.
Our calculations show that being presented in the $\Lambda S$ coupling scheme, the spin-orbit mixed $H$ state of ThO has the main contribution (more than 95\%) from the $^3\Delta_1$ configurations.
Therefore, due to the identity  $\langle\Psi_{^3\Delta_1}|S^e_+ |\Psi_{n0^\pm}\rangle \equiv 0$ for pure $\Lambda S$ state, the inequality $|\langle\Psi_{H^3\Delta_1}|S^e_+ |\Psi_{n0^\pm}\rangle| \ll 1 $ holds with a good accuracy. This inequality gives sufficient condition for $\Delta g$ to be determined by the energy splitting between the top and bottom levels of the $\Omega-$doublet \cite{Petrov:11, Lee:13a}. The rotational analysis given in \cite{Edvinsson:84} for $P(\Omega=0)- H^3\Delta_1$ and $O(\Omega=0)- H^3\Delta_1$ bands have shown that the $\Omega-$doublet spacing ($\Delta=|E(J=1^-) - E(J=1^+)|$) in $H^3\Delta_1$ is $350-470$~kHz. At the moment, the parity assignment for electronic states $P$ and $O$ is yet unclear. According to \cite{Edvinsson:90} $P$ is $0^-$ and $O$ is $0^+$. The latter, as can be shown, indicate that $e$ states are the top levels whereas the $f$ states are the bottom levels of the $\Omega$-doublets for $H^3\Delta_1$. 
The latter-day microwave spectroscopy confirms our conclusion about the levels ordering and finally gives $E(J=1^-) - E(J=1^+)=\rm{362}\pm 10$kHz \cite{ACME:13}. 
%
In Fig.~\ref{gfgecross} the calculated $g$ factors for the $J = 1$ levels of ThO $H^3\Delta_1$ state are given as functions of the laboratory electric field.
 The lowest value, $\Delta g = 2.7\times10^{-6}$, is attained at the electric field $4.4$ V/cm.
Note that the molecule is completely polarized at the electric field larger than 3~V/cm.

\section{Conclusion}
A number of parameters (\Eeff~ and $W_{T,P}$) that are required to interpret experimental measurements on the $H^3\Delta_1$ state of ThO molecule in terms of fundamental quantities are calculated.
Though the previous estimation of \Eeff~ made in Ref. \cite{Meyer:08} is only 25\% more than our final value, the good agreement can be rather considered as ``fortunate'' since the accuracy of the semiempirical estimates for the systems like ThO having a very complicated electronic structure is severely limited, see \cite{Skripnikov:09}. In turn, the reliable ab initio calculation of ThO is on the threshold of current possibilities of computational methods and we estimate the accuracy of our calculation of Eeff by 15\% only. Nevertheless, even such accuracy is important to establish a reliable eEDM estimate in the ongoing ThO experiment compared to the measured upper bounds on eEDM in Tl~\cite{Regan:02} and YbF~\cite{Hudson:11a}  experiments.


\section{Acknowledgement}
We are grateful to Professor DeMille for useful discussions and remarks.
Also we are grateful to N.S.~Mosyagin for providing us with the GRECP for Th \cite{Mosyagin:13a}.
This work is supported by the SPbU Fundamental Science Research grant from Federal budget No.~0.38.652.2013 and RFBR Grant No.~13-02-01406. L.S.\ is also grateful to the Dmitry Zimin ``Dynasty'' Foundation. The molecular calculations were performed at the Supercomputer ``Lomonosov''.


\end{document}